# Generation of Spatially Coherent Light at Extreme Ultraviolet Wavelengths

Randy A. Bartels,[1] Ariel Paul,[1] Hans Green,[1] Henry C. Kapteyn,[1] Margaret M. Murnane,[1] Sterling Backus,[1] Ivan P. Christov,[2] Yanwei Liu,[3,4] David Attwood,[3,4] and Chris Jacobsen[5]

1. Department of Physics and JILA, University of Colorado and National Institute of Standards and Technology, Boulder, CO 80309-0440
2. Dept. of Physics, Sofia University, Sofia, Bulgaria
3. Center for X-ray Optics, Lawrence Berkeley National Laboratory, Berkeley, CA 94720
4. Applied Science and Technology, University of California, Berkeley, CA 94720
5. Department of Physics and Astronomy, SUNY at Stony Brook, Stony Brook, NY 11794

We present spatial coherence measurements of extreme-ultraviolet light generated using the process of high-harmonic upconversion of a femtosecond laser. Using a phase-matched hollow-fiber geometry, the generated beam is found to exhibit essentially full spatial coherence. The coherence of this laser-like EUV source is demonstrated by recording Gabor holograms of small objects. This work demonstrates the capability to do EUV holography using a tabletop experimental setup. Such an EUV source, with low divergence and high spatial coherence, can be used for experiments such as high-precision metrology, inspection of optical components for EUV lithography (*1*), and for microscopy and holography (2) with nanometer resolution. Furthermore, the short time duration of the EUV radiation (a few femtoseconds) will enable EUV microscopy and holography to be performed with ultrahigh time resolution.

One of the fundamental properties of a laser is the ability to produce spatially coherent beams, phase coherent across the radiation field. This is accomplished in the visible and ultraviolet regions of the spectrum through the use of an optical resonator which permits the growth of only a single transverse "TEM$_{00}$" mode. At shorter wavelengths, in the EUV and soft x-ray regions of the spectrum, optical resonators are of limited use due to the finite lifetime of the gain medium, limited reflectivity of available mirrors, and mirror damage. Short-wavelength light sources such as electron impact sources, synchrotron sources (3), x-ray lasers (4, 5), and free-electron lasers (6) to date generate only partially-coherent light. Pinhole spatial filtering has been used to achieve spatial coherence from otherwise incoherent sources, such as undulator radiation, but generally at a significant loss of available photon flux, and without imposing phase coherence across the radiation field. In this paper we discuss the use of high harmonic generation (HHG) which is compact in nature, generates phase coherent radiation across the full field, and does so with femtosecond pulses ideal for the dynamical studies of many systems.

High harmonic generation (HHG) is a useful way of generating coherent light in the ultraviolet and extreme ultraviolet (EUV) regions of the spectrum. In HHG (*7, 8*), pulses of short-wavelength light with extremely short duration (*9, 10*) can be produced by focusing a high-intensity femtosecond laser into a gas. Odd harmonics of the exciting laser frequency (i.e. 3ω, 5ω, etc.) are produced in a directed, narrow-divergence beam, with photon energies that can extend up to > 500eV (corresponding to harmonic orders > 300). This process thus coherently up-shifts a femtosecond pulse from the visible into the EUV region of the spectrum. However, experiments to date have shown that the HHG radiation has only partial spatial coherence, and thus does not retain the full coherence of the fundamental driving beam (*11-13*).

From a classical viewpoint, high-harmonic generation driven by coherent light from a laser is itself a coherent process since the nonlinear-optical response of the atoms involved in the upconversion process is fully coherent and deterministic. Thus, one might expect that the HHG process would generate fully coherent light. However, measurements to date have shown that mechanisms such as plasma refraction, as well as a complex spatial and temporal phase of the generated light that results from the quantum nature of the HHG process, limit the coherence of HHG-generated EUV light. The HHG process is unique as a coherent optical process in that high-harmonics are generated by atoms during the process of ionization - electrons ionized by the strong field created by an ultrashort laser can "recollide" with their parent ion as they begin to oscillate in the laser field (*14, 15*). This recollision process results in the EUV emission; however, it also dramatically and dynamically changes the index of refraction of the medium. This large, time-varying index has been identified as the reason why early experiments have measured only partial spatial coherence (*11-13*). Degraded coherence also results from the fact that emission at any particular wavelength can result from many electron recollision trajectories, creating a complex and spatially-varying multimode wave front (*12*). Ionization occurs twice each cycle of the optical driving field, as the magnitude of the field reaches its peak. However, for each particular harmonic order, two separate electron trajectories (corresponding to slightly differing ionization and recollision times within the optical cycle), generate the same photon energy (*15, 16*). Moreover, for HHG excited by relatively long laser pulses, many different optical cycles contribute to a given harmonic order. All these effects can reduce the coherence of the source. Although the proper selection of experimental conditions, such as the position of the focus with relation to the position of the nonlinear medium, can optimize phase-matching in the forward direction and partially mitigate these effects (*13*), simple optimization has not succeeded in fully regaining the coherence of the source.

In past work, we demonstrated that the HHG process can be phase-matched over a long interaction region using a hollow core fiber (*17-19*). This geometry increases the conversion efficiency of light into the EUV by up to two orders of magnitude over what would be possible using similar pulse energies in a free-space focus configuration. Furthermore, the quasi-plane wave interaction in the hollow fiber and the long propagation distance in the nonlinear medium select a single "recollision" trajectory and serves to improve the temporal coherence. Here, we show that this extended phase matching also significantly improves both the beam mode quality and the spatial coherence, by phase-matching only the emission of individual atoms that contribute to a fully-coherent, forward-directed beam of short-wavelength light.

In this work, light from a high repetition rate (5 kHz, ~0.8 mJ/pulse) Ti:sapphire laser system (*20*) operating at 760 nm with a pulse duration of 25 fs, was focused into a 10 cm long, 150 µm diameter, hollow-core fiber filled with Argon gas (Fig. 1). The EUV radiation is phase-matched at a pressure of 29 Torr, resulting in emission of ≈ 3-5 harmonics centered around 31 eV photon energy (harmonic orders 17 – 23). The pump pulse propagates predominantly in the $EH_{11}$ mode of the hollow core fiber, while the high-harmonic generation is restricted to the central, most intense portion of the fundamental mode. Note that a 0.55 µm thick Aluminum filter is used to remove the fundamental laser light, and is immediately followed by the object to be imaged. In an image of the EUV beam 95 cm after the exit of the hollow core fiber (Fig. 1), the diameter of the EUV beam is 1 mm at the $1/e^2$ point, with a slight ellipticity (~1.3) due to imperfections in the hollow-fiber shape. The beam divergence of < 1 mrad is consistent with a diffraction-limited source size of 40 µm diameter. Using a vacuum photodiode, we measure a photon flux of approximately $2 \times 10^{12}$ photons/second.

The spatial coherence of the EUV light was measured using the double-pinhole (or double slit) interference technique (*21*). The depth of modulation of the fringes generated after passing a beam through a pinhole pair depends on the correlation between the local phase of the wave front of the beam at the two points where it is sampled by the pinhole pair. If the phase difference between the two points is constant and deterministic (and therefore completely correlated), the fringe depth will be unity. However, if there are random variations in the phase between the two points, the fringe contrast will be degraded due to implicit detector averaging. The fringe visibility was measured across the width of the EUV beam by sampling it with pinhole pairs separated by between 142 and 779 µm. We used apertures (National Aperture, Inc) fabricated with 20 or 50 µm diameter pinhole pairs, placed 95 cm from the exit of the fiber. An EUV Charge-coupled

Device (CCD, Andor Inc.) camera placed 2.85 m from the pinholes captured the diffracted image. High dynamic range images were possible by using a CCD integration time of between 20 and 240 s (100,000 – 1,200,000 laser shots). Integration over a large number of shots demonstrates both the high spatial coherence and the long-term wavefront stability of the EUV beam.

In the measured diffraction patterns for a set of pinhole pairs (Fig. 2), the fringe visibility varies across the pattern. Coherence measurements are usually performed using quasi-monochromatic radiation, in which case the visibility is constant over the entire pattern. When the incident radiation is broadband (in our case consisting of several EUV harmonics), the modulation depth at the center of the fringe pattern (equidistant from the two pinholes) corresponds to the fringe visibility. Analysis of the full modulation depth of the interference pattern over the entire field can yield information about the incident spectrum. The diffraction pattern produced by a uniformly-illuminated pinhole pair can be written as –

$$I(x) = 2I^{(0)}(x)\left[1 + \gamma_{12}(x)\cos\left(2\pi \frac{d}{\lambda_0 z}x\right)\right], \qquad (1)$$

where $I^{(0)}(x)$ is the Airy distribution due to diffraction through a pinhole of width $\delta$, $d$ is the pinhole separation, $z$ is the distance from the pinhole pair to the observation plane, $\lambda_0$ is the central wavelength, and $\gamma_{12}$ is the degree of mutual coherence defined as the magnitude of the complex degree of mutual coherence ($\boldsymbol{\gamma}_{12}(x) = \gamma_{12}(x)\exp\left(-i2\pi \frac{d}{\lambda_0 z}x\right)$). The Fourier transform of equation 1 is written as –

$$\Im\left\{I\left(\tau = \frac{d}{zc}x\right)\right\} = 2T(v) \otimes \left\{\delta(v) + \frac{1}{2}\hat{S}(v)\mu_{12}(v) \otimes \left[\delta\left(v - \frac{zc}{d}f_0\right) + \delta\left(v + \frac{zc}{d}f_0\right)\right]\right\} \qquad (2)$$

where $\otimes$ is the convolution operator, $T(\nu) = \Im\left\{I^{(0)}\left[\left(\frac{d}{zc}\right)x\right]\right\}$ is a "dc" spike, $\delta(\nu)$ is the Dirac delta function, $f_0 = \frac{1}{\lambda_0}\frac{d}{z}$ is the carrier spatial frequency, and $\hat{S}(\nu)\mu_{12}(\nu) = \Im\{\gamma_{12}(\tau)\}$ (where $\hat{S}(\nu)$ is the power spectrum normalized such that $\int_0^\infty \hat{S}(\nu)d\nu = 1$) (*22, 23*). Equation 2 contains three terms: a "dc", $T(\nu)$, term due to diffraction through a single pinhole, and two "sideband" terms that contain both spectral information and fringe visibility. Each sideband term contains the product of the fringe visibility as a function of frequency and the spectrum of the incident field convolved with the dc term.

In the case of quasi-monochromatic radiation, the spatial coherence factor, $\mu_{12}$, is simply twice the height of one of the sideband terms after the maximum value of the dc spike has been normalized to unity. More generally, we can sum the integral of the sidebands and divide by the integral of the dc term, resulting in the following expression –

$$\tilde{\mu}_{12} = \frac{\int T(\nu) \otimes \hat{S}(\nu - \nu_0)\mu_{12}(\nu - \nu_0)d\nu + \int T(\nu) \otimes \hat{S}(\nu + \nu_0)\mu_{12}(\nu + \nu_0)d\nu}{\int 2T(\nu)d\nu}. \quad (3)$$

Note that this expression defines an average fringe visibility weighted by the spectral intensity ($\tilde{\mu}_{12} = \int \hat{S}(\nu)\mu_{12}(\nu)d\nu$) (*21, 23*). For monochromatic light, the spectrum is a delta function and the fringe visibility at the central frequency is obtained directly. In our experiment, the broad bandwidth EUV spectrum (consisting of harmonic orders 17 through 23) means that we obtain a spectral average of the fringe visibility. However the extremely high observed coherence means any variation across the spectrum is minimal.

The EUV beam is sampled at 14%, 24%, 29%, 38%, 58%, and 78% of the beam diameter using pinhole pair separations of 142, 242, 292, 384, 574, and 779 μm respectively (separations verified by an SEM). Sample data is shown in Fig. 2. Under the far-field conditions of these measurements, Eqn. 3 is valid for all but the two greatest pinhole separations. We simply Fourier transform the data, identify the sidebands, and integrate to obtain the average spatial coherence. For the 779 μm pinhole pair (Fig. 2D), the interference pattern contains two circular Airy distribution patterns with fringes. The two Airy distributions are separated by ~3.2 mm, compared with the pinhole separation of 779 μm. This is due to the fact that the pinholes are sampling the curvature of the EUV phase front, and the local tilt is larger than the divergence due to diffraction. Here, Eqn. 1 does not apply, since the two beams do not equally illuminate each point on the CCD. Nevertheless, in the central region where the two Airy patterns have equal intensity, the observed fringe visibility can be used as a minimum measure of $\tilde{\mu}_{12}$. For all other pinhole separations, the two methods agree. Fig. 3 shows the magnitude of the complex coherence function as a function of pinhole separation. It is evident from this plot that we maintain unity spatial coherence over most of the EUV beam.

Our entire set-up, including the femtosecond laser system, x-ray generation cell, imaging set-up and x-ray CCD camera, occupies 100 cm x 350 cm of optical table space. This compact, coherent, laser-like, source of tabletop EUV radiation is extremely useful for applications such as high-resolution coherent imaging. For example, techniques such as Gabor holography (*24*) can measure both the amplitude and phase transmittance (optical density) of objects with small amounts of absorption, or can image small objects that strongly absorb or scatter radiation. Gabor holography also is extremely simple, requiring only coherent illumination of an object, with no optical system, as shown in Fig. 1. This is advantageous in spectral regions such as the EUV where few optical elements are available, and those that are have a maximum reflectivity of ~ 70 % for a few limited wavelengths. The resolution of Gabor holography using a collimated reference beam is approximately equal to the resolution of the detector (*25*). However, if a diverging beam is used, such as in our setup, the resolution of the hologram is increased by the geometric projection magnification. We demonstrate the utility of our coherent EUV radiation by recording a Gabor hologram of a near field scanning optical microscopy (NSOM) tip and a ~ 10 micron diameter water jet. We note that although this is the first time to our knowledge that a table-top EUV source has been used to record and reconstruct holographic images, past work has demonstrated simple interferometry using small-scale EUV sources (*26, 27*), as well as holography using EUV and soft x-ray light from spatially-filtered large-scale light sources (*28, 29*).

The NSOM tip (Fig. 4A) was placed 78.5 cm away from the exit of the hollow core fiber where the EUV beam is generated. The Gabor hologram was then recorded with the CCD camera, at a distance of 2.15 m from the NSOM tip. Under these conditions, the geometric magnification of 3.8, coupled with the 26-micron resolution of our detector, gives rise to a resolution of 6.8 micron. (Fig. 4B) In the case of the water jet, the object was placed 95.5 cm away from the exit of the hollow core fiber, while the hologram was recorded at a distance of 2.17 m. (Fig. 4D) This geometry gives rise to a magnification of 3.3 and yields a resolution of 7.9 microns. The holograms are reconstructed (Fig. 4C,4E) using a standard numerical technique (*30*).

The resolution of the Gabor holography in this setup can be improved by increasing the detector resolution, using a larger geometric magnification, or by optical magnification using an EUV optic. For example, by increasing the source-detector to source-object distance ratio to ~30x, and by using a commercially-available CCD with 13 µm pixels, a resolution of <500 nm could be obtained. The use of a simple magnifying optic could increase the effective detector resolution further, ultimately allowing resolutions approaching λ~30 nm. Further improvements should allow this source to be used for applications in high-precision metrology, inspection and metrology for EUV lithography masks, coatings, and optics, plasma imaging, and for microscopy and holography with nanometer resolution. The time duration of the HHG light pulses is less than 10 femtoseconds, so that complimentary experiments can be contemplated involving stroboscopic holography with a time resolution of ≈ 10 fsec, limited only by the object illumination geometry.

# FIGURE CAPTIONS

**Figure 1:** Experimental set-up and beam profile of the EUV light measured 95 cm from the exit of the fiber.

**Figure 2:** Interferogram images of the EUV beam diffracted by pinhole pairs of various separations, together with lineouts of the images. (A) = 142 µm, (B) = 242 µm, (C) = 384 µm, (D) = 779 µm separation.

**Figure 3:** Spatial coherence of the EUV beam as a function of pinhole separation.

**Figure 4:** Reconstruction of Gabor holograms of two objects generated with our EUV beam. The first hologram is the tapered fiber of a Near-field Scanning Optical Microscope (NSOM) tip (A), its reconstructed image (B), and a picture of the NSOM tip taken with a conventional visible microscope (C) for reference. The second hologram is of a ~ 10 micron diameter water jet (D), and its reconstruction (E).

**References and Notes**

1. K. A. Goldberg *et al.*, *J. Vac. Sci. Technol.* **16**, 3435-3439 (1998).
2. E. Leith, J. Roth, *Appl. Opt.* **16**, 2565-2567 (1977).
3. D. Attwood, *Soft x-rays and Extreme Ultraviolet Radiation,* Cambridge University Press: New York, NY (1999).
4. D. L. Matthews *et al.*, *Phys. Rev. Lett.* **54**, 110-3 (1985).
5. C. Macchietto, B. Benware, J. Rocca, *Opt. Lett.* **24**, 1115-1117 (1999).
6. J. Andruszkow *et al.*, *Phys. Rev. Lett.* **85**, 3825-3829 (Oct 30, 2000).
7. A. McPherson *et al.*, *J. Opt. Soc. Am. B* **4**, 595-601 (1987).
8. A. L'Huillier, K. J. Schafer, K. C. Kulander, *Phys. Rev. Lett.* **66**, 2200 (1991).
9. Z. Chang, K. Kim, H. Wang, H. Kapteyn, M. Murnane, paper presented at the Conference on Laser and Electro-optics: CLEO '99, Baltimore, MD 1999.
10. M. Drescher *et al.*, *Science* **291**, 1923-1927 (2001).
11. T. Ditmire *et al.*, *Phys. Rev. Lett.* **77**, 4756-9 (1996).
12. P. Salieres, A. L'Huillier, M. Lewenstein, *Phys. Rev. Lett.* **74**, 3776 (1995).
13. P. Salieres, T. Ditmire, M. Perry, A. L'Huillier, M. Lewenstein, *J. Phys. B* **29**, 4771-4786 (1996).
14. K. C. Kulander, K. J. Schafer, J. L. Krause, in *Super-intense laser-atom physics* B. Piraux, A. L'Huillier, K. Rzazewski, Eds. (Plenum Press, New York, 1993), vol. 316, pp. 95-110.
15. M. Lewenstein, P. Balcou, M. Y. Ivanov, A. L'Huillier, P. B. Corkum, *Phys. Rev. A* **49**, 2117-2132 (1994).
16. M. Lewenstein, P. Salieres, A. L'Huillier, *Phys. Rev. A* **52**, 4747-4754 (1995).
17. A. Rundquist *et al.*, *Science* **280**, 1412-1415 (1998).
18. C. G. Durfee *et al.*, *Phys. Rev. Lett.* **83**, 2187-2190 (1999).
19. R. Bartels *et al.*, *Nature* **406**, 164-166 (2000).
20. S. Backus *et al.*, *Opt. Lett.* **26**, 465-467 (2001).
21. B. Thompson, E. Wolf, *J. Opt. Soc. Am.* **46**, 895 (1957).
22. A. T. Friberg, E. Wolf, *Opt. Lett.* **20**, 623-625 (Mar 15, 1995).
23. R. A. Bartels, A. Paul, M. M. Murnane, H. C. Kapteyn, S. Backus, *Opt. Lett.* **In press** (2002).
24. D. Gabor, *Nature* **161**, 777-778 (1948).
25. S. Lindaas, H. Howells, C. Jacobsen, A. Kalinovsky, *J. Opt. Soc. Am. A* **13**, 1788-1800 (Sep, 1996).
26. M. Marconi, C. Moreno, J. Rocca, V. Shlyaptsev, A. Osterheld, *Phys. Rev. E* **62**, 7209-7218 (2000).
27. D. Descamps *et al.*, *Opt. Lett.* **25**, 135-137 (2000).
28. L. Da Silva *et al.*, *Rev. Sci. Instrum.* **66**, 574-578 (1995).
29. S. H. Lee *et al.*, *Appl. Opt.* **40**, 2655-2661 (June 1, 2001).
30. U. Schnars, W. P. O. Juptner, *Appl. Opt.* **33**, 4373-4377 (Jul 10, 1994).
31. This work was supported by the National Science Foundation Grant No. ECS-0099886 and the Department of Energy High Energy Density Science Grants Program. This work made use of facilities funded by the W.M. Keck Foundation.

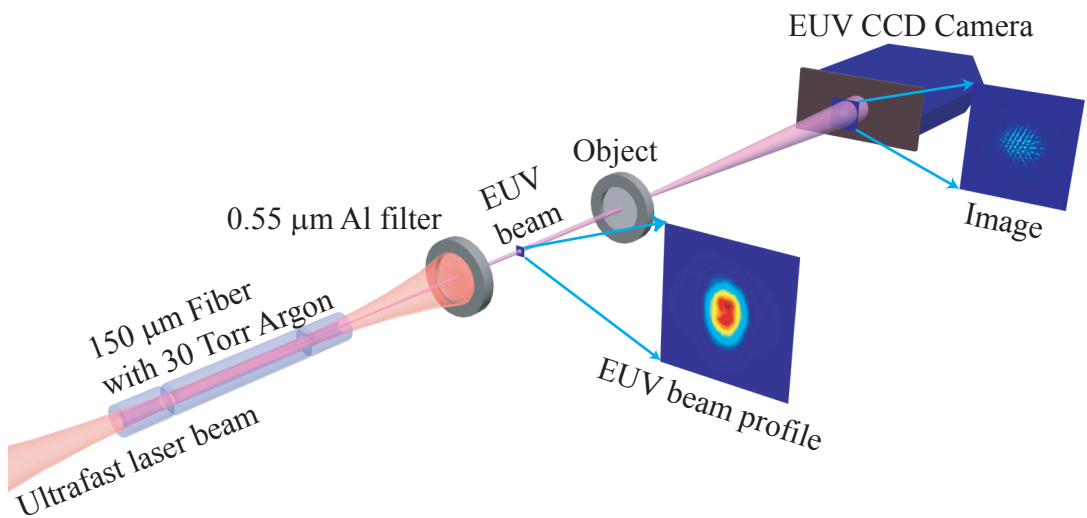

Randy Bartels, *et al.*; Figure 1

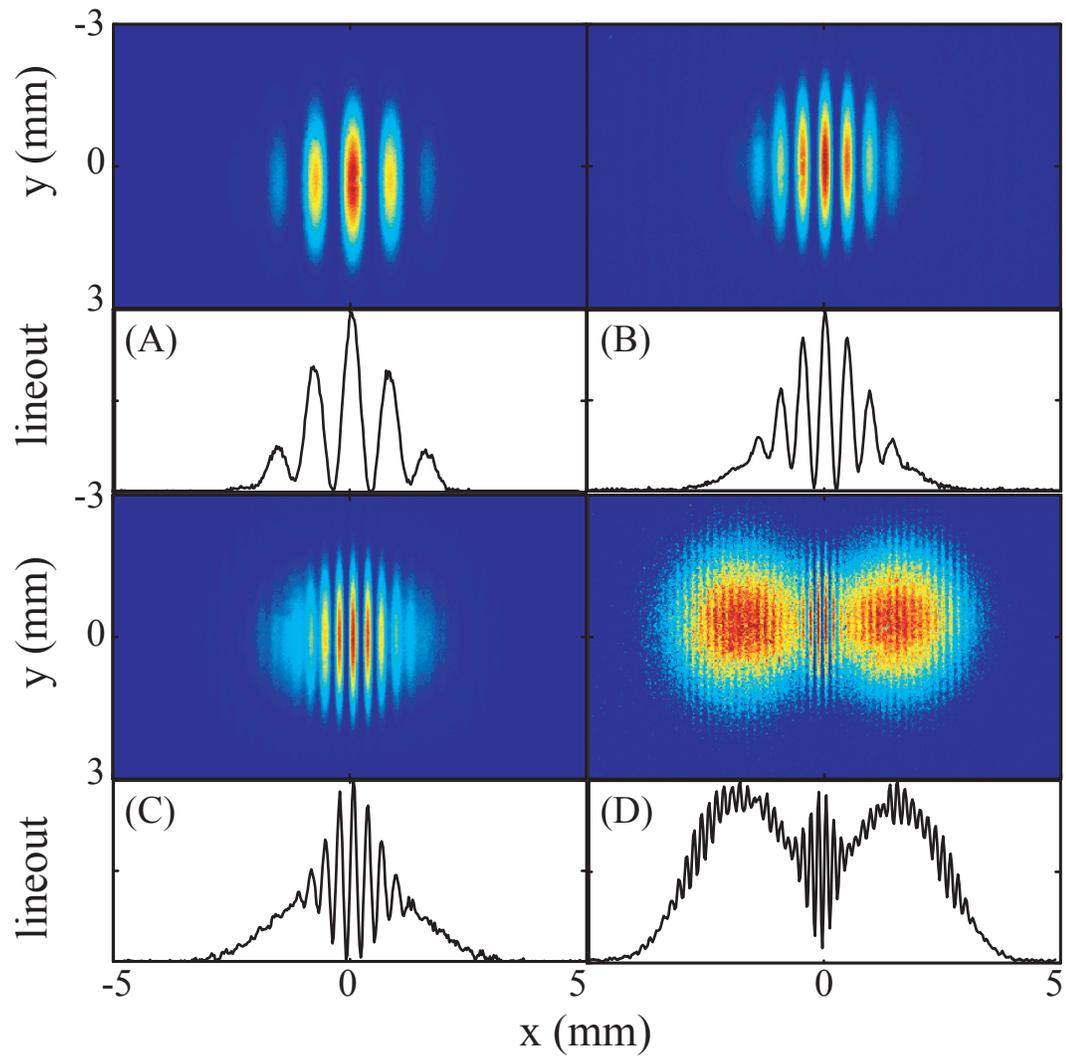

Randy Bartels, *et al.*; Figure 2

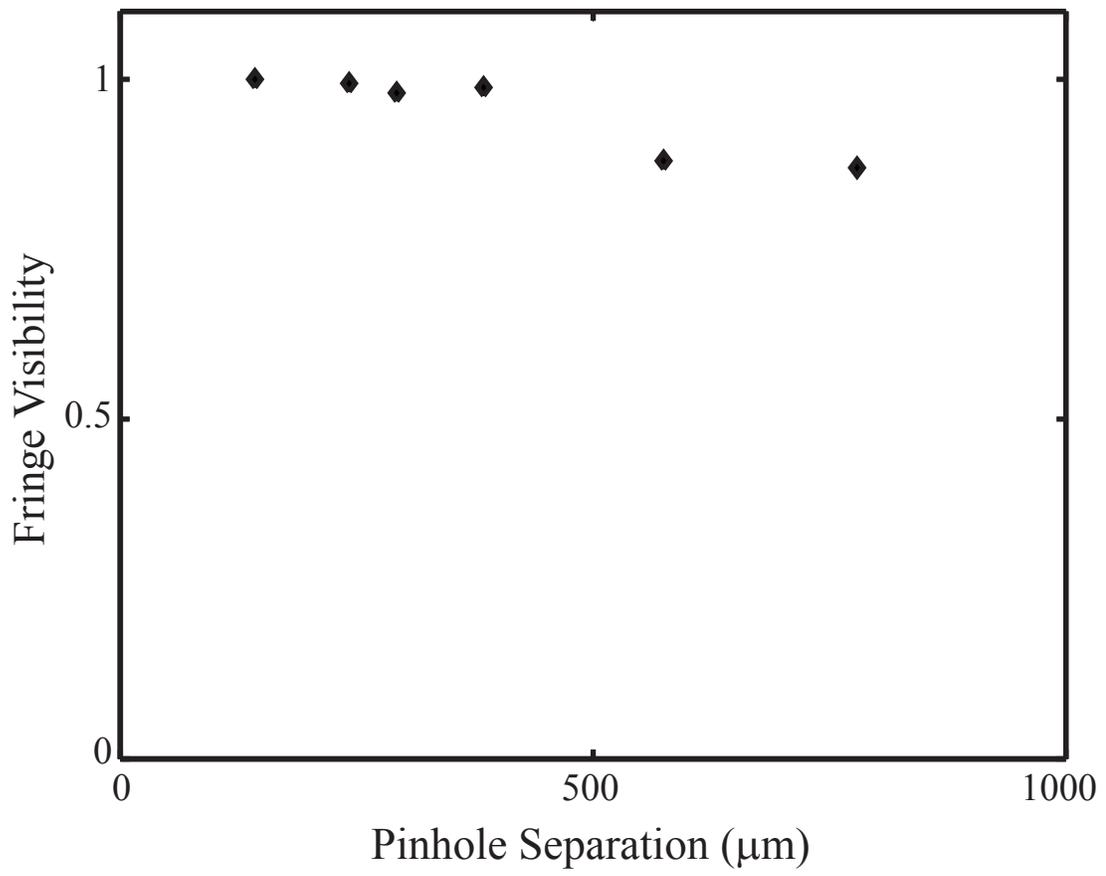

Randy Bartels, *et al.*; Figure 3

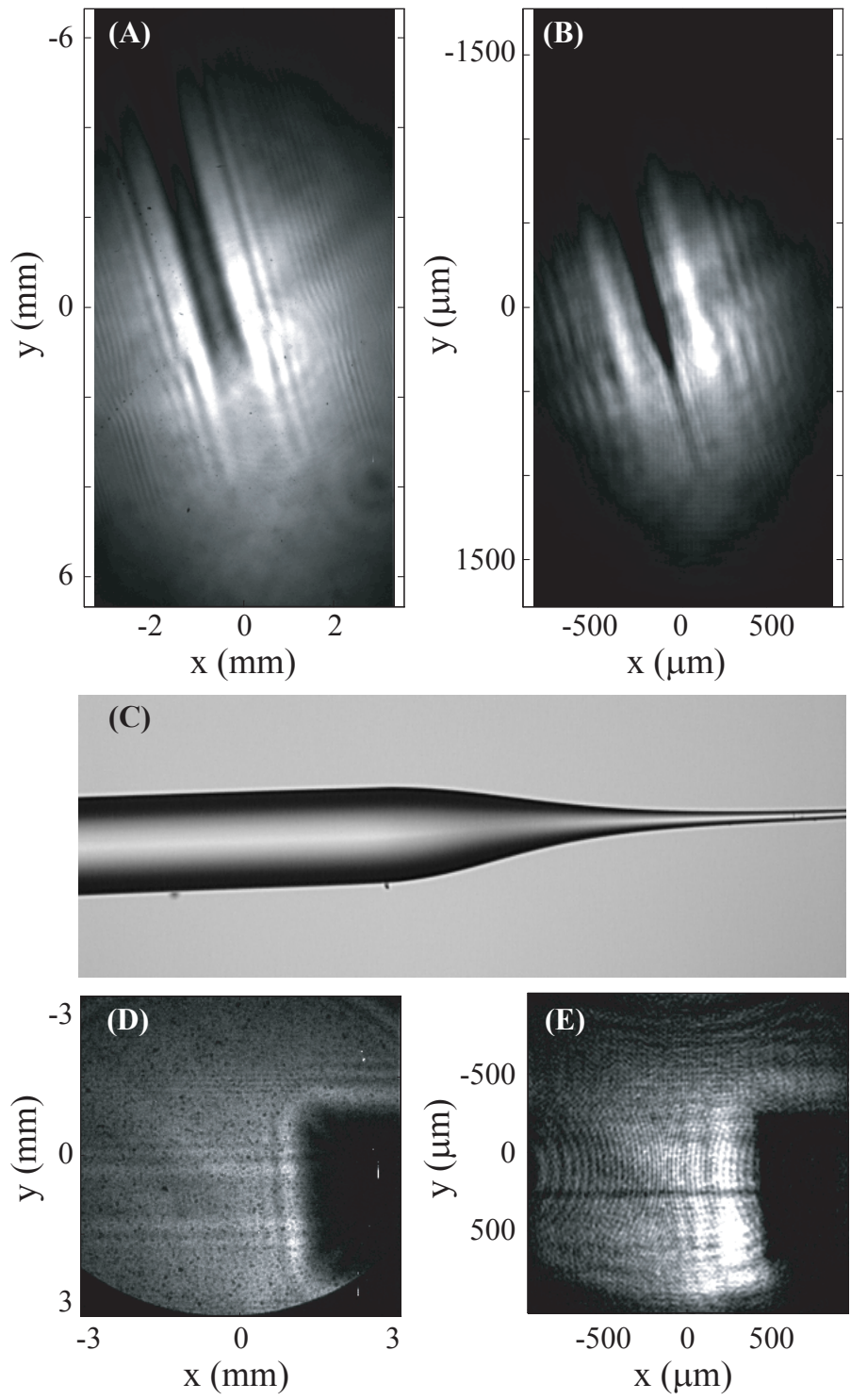

Randy Bartels, *et al.*; Figure 4